\documentstyle[12pt]{article}
\title{Behaviour of superconductivity energetic characteristics in
electron-doped cuprates. A simple model}

\author{N. Kristoffel$\sp{*,\dagger}$, P. Rubin\footnote{Institute of Physics, University of Tartu, Riia 142, 51014
Tartu, Estonia \newline $\dagger$ Institute of Theoretical Physics,
University of Tartu, T\"ahe 4, 51010 Tartu, Estonia}}

\date{}
\begin{document}
\maketitle
\begin{minipage}{12cm}

 A simple model to describe the energetic phase diagram of
electron-doped cuprate superconductor is developed. Interband
pairing operates between the UHB and the defect states created by
doping and supplied by both extincting HB-s. Two defect subbands
correspond to the ($\pi ,0$) and ($\pi /2,\pi /2$) momentum regions.
Extended doping quenches the bare normal state gaps (pseudogaps).
Maximal transition temperature corresponds to overlapping bands
ensemble intersected by the chemical potential. Illustrative results
for $T_c$, pseudo- and superconducting gaps are calculated on the
whole doping scale. Major characteristics features on the phase
diagram are reproduced. Anticipated manifestation of gaps doping
dynamics is discussed.
\end{minipage}

\section{Introduction}
\label{intro} Electron-doped high-temperature cuprate
superconductors are represented by materials like
A$_{2-x}$Ce$_x$CuO$_4$ with A = La; Nd; Pa; Sm; Eu. These systems
have been investigated [1-7] not so widely as the hole-doped
counterparts of them. In both cases the doping process does not only
metallizes the basic compound but prepares also a modified
playground for the realization of the superconductivity. The
electron spectrum of such strongly correlated systems is nonrigid
under doping. This leads to the reconstruction of the Fermi surface
or its fragments [8-11]. The investigation of electron-doped
materials is essential for covering the whole active part of the
energetic spectrum of a cuprate superconductor. It originates from
the spectrum of a half-filled charge-transfer insulator and
undergoes remarkable reorganization due to doping. In the case of
hole doping the region near the top of the mainly oxygen band
between the splitted Cu Hubbard components participates essentially
in the superconductivity. In the case of electron doping the Hubbard
structure is expected to be immediately encompassed with essential
events near the bottom of the upper Hubbard band (UHB). For both
types of doped cuprates a "defect band'' near the Fermi level is
created under the perturbation exerted by the introduced extra
carriers [8-11]. For hole doping this new band borns near the top of
the oxygen band (citations in [12,13]). Extended doping brings it
into overlap with the itinerant oxygen band [13]. A phase separation
appears [14] which reorganizes the crystalline background in
hole-rich and hole-poor (defect and itinerant correspondingly)
regions. In the case of electron doping the defect band will be
built up under the bottom of the UHB [11], see also the theoretical
papers [15-18]. A phase separation seemingly does not accompany the
collapse of the Hubbard band spectrum with extended doping. Both
components of the latter donate the states to built up the new band
in this process.

The most remarkable difference between electron- and hole doped
cuprate superconductors consists in the essentially wider
antiferromagnetic (AF) region in the former case. It overlaps
partly the restricted superconductivity region with moderate $T_c$
values. The pseudogap feature is currently known also for the
electron doped systems [3-5,16,19,20]. There are some indications
of presence of two pseudogaps connected with the quenching of the
AF order by extended doping [4,20]. The pseudogap vanishes near
the optimal doping distinguishable on the bell-like dependence of
$T_c$. Decisive data on measured superconducting gaps in electron
doped cuprates seem to be rare.

The pairing mechanism in cuprate high-temperature superconductors
remains elusive. For this reason under a plenty of approaches a
simple model basing on gross features known experimentally has
been developed for description of hole doped cuprates [10,21,22].
On such a way one can presumably prove the general direction for
elaborating a throughout theory and understand the nature of the
pairing in cuprates. In the model mentioned the pairing
interaction is supposed to be of interband nature between the
itinerant and defect band states. There are numerous appointments
that cuprate superconductors are multigap and multiband systems
being recently supported by the direct measurement of two
superconducting gaps in the LaSrCuO system [23].

The multiband superconductivity incorporating interband pairing is
known for a long time [24,25]. It serves the simplest way to
obtain high $T_c$ values in the dispositions where several bands
cross the Fermi surface and quantum resonance effects appear (for
review see [26-28]). Recently the interest to multiband models in
application to cuprates has essentially grown [32,34] stimulated
also by the discovery of the multiband superconductor MgB$_2$. The
model [19,21,22] describes self-consistently the behaviour of
energetic [12,21,22], thermodynamic and other properties [29-31]
of hole doped cuprates on the whole doping scale in qualitative
agreement with the observations. The conception of the
multicomponent nature of the cuprate superconductivity [32] is
supported by this.

In the present contribution an attempt to develop an analogous
model for the electron doped suprate superconductors is made.

\section{The Model}
The total number of states in the system including LHB, UHB and
the defect band components supposed to be actual for the electron
doped cuprate is normalized to one. The doped electron
concentration for one Cu site and one spin will be denoted as $c$.
According to [11] the UHB ($\beta$) and LHB ($\gamma$) weight of
states under electron doping is ($1/2-c$). Note that this
restricts the doping concentrations in our model to $c<1/2$. The
states exhausted from the Hubbard bands concentrate in the defect
(midgap) band with the weight 2c. The "itinerant'' band densities
are $\rho_{\beta ,\gamma}=(1/2-c)/\Delta d$, where $\Delta d$ is
the corresponding band width (2D case of CuO$_2$ plane as
superconductivity playground). Experimental data show that the
functioning of momentum space regions around the $(\pi ,0)$ and
$(\frac{\pi}{2}, \frac{\pi}{2})$ type points in the Brillouin zone
must be differentiated near the Fermi energy, at least for
underdoping. Therefore we introduce two defect subbands of weight
$c$ displaced in energy by $d$. The upper ($\alpha 1$) corresponds
to the $(\pi ,0)$ neighbourhood, as also the $\beta$-band bottom
(energy $d_1$). The lower subband ($\alpha 2$) corresponds to the
$(\frac{\pi}{2}, \frac{\pi}{2})$ neighbourhood. The $\alpha 2$
band bottom is taken as energy zero.

The doping dependence of the pseudogap type excitations
[4,6,19,20] and the build up of a common Fermi liquid at extended
dopings suggest that the $\alpha 1$ and $\alpha 2$ top energies
must evolve towards the $\beta$-band bottom. This is described by
writing the corresponding energies as $(d+\alpha c\sp 2)$ and
$\alpha c\sp 2$. The quadratic dependence on concentration $c$ has
been suggested by the form of the pseudogap curves in [4,20]. The
defect subbands densities $\rho_{\alpha}(1,2)=(\alpha c)\sp{-1}$
rise with doping and can be considered as determining a "flat
band''.

The $(\alpha 1)-\beta$ gap closes with extended doping at the first
critical point of the phase diagram with the concentration defined by
$c_0\sp 2=(d_1-d)\alpha\sp{-1}$. For $c\leq c_0$ the chemical potential
\begin{equation}
\mu_1=d+\alpha c\sp 2
\end{equation}
remains connected with the upper defect subband $\alpha 1$ (the
$\alpha 2$ component is filled). This agrees with the observation
that at low dopings the electrons occupy the ($\pi ,0$)
neighborhood which forms the Fermi surface pockets [5-7]. For
$c\geq c_0$ the $\beta$ and $\alpha 1$ bands overlap and become
populated. Then
\begin{equation}
\mu_2=(d+\alpha c\sp 2+d_1\alpha c\rho_{\beta})(1+\alpha c\rho_{\beta})\sp{-1}\, .
\end{equation}

Two distinct (steep and flat) dispersions near ($\pi ,0$) have been
measured in [6].

A further critical point $c_x$ appears when $\mu_2(c_x)=\alpha
c_x\sp 2$ and all the band components become intersected by the
chemical potential. Such disposition is optimal for producing
maximal transition temperatures by the interband pairing
mechanism. For $c>c_x$
\begin{equation}
\mu_3=(d + 2\alpha c\sp 2 + d_1\alpha c\rho_{\beta})(2 + \alpha
c\rho_{\beta})\sp{-1} \; .
\end{equation}

Accordingly a new spectral intensity around
($\frac{\pi}{2},\frac{\pi}{2}$) emerges, cf. [6,7]. The wider
$\beta$-band contributes in the effective doping range to the
formation of an electron-like Fermi surface. The chemical
potential rises with electron doping in our model as stated in
[2].

The central supposition of our approach is that the basic pairing
channel consists in the pair transfer interaction [26] between the
itinerant $\beta$ and defect $\alpha 1,2$ band states.

\section{Superconductivity characteristics}
We describe our superconducting system by the Hamiltonian
$$
H=\sum_{\sigma ,\vec{k},s}\epsilon_{\sigma}(\vec{k}) a\sp
+_{\sigma ,\vec{k},s}a_{\sigma ,\vec{k},s}+ \Delta_{\beta}
\sum_{\vec{k}} [a_{\beta\vec{k}\uparrow}a_{\beta
-\vec{k}\downarrow}+ a\sp +_{\beta -\vec{k}\downarrow}a\sp
+_{\beta \vec{k}\uparrow}]
$$
\begin{equation}
-\Delta_{\alpha}\sum_{\vec{k},\alpha}{}\sp{1,2}
[a_{\alpha\vec{k}\uparrow}a_{\alpha -\vec{k}\downarrow}+ a\sp
+_{\alpha -\vec{k}\downarrow}a\sp +_{\alpha \vec{k}\uparrow}]\; .
\end{equation}

Here $\epsilon_{\sigma}=\xi_{\sigma}-\mu$, $\sigma =\alpha
,\beta$, $\sum_{\alpha}{}\sp{1,2}$ means the integration with the
density ($\rho_{\gamma}$) in the corresponding energy intervals
for the defect bands. Usual designations for electron operators
and spins ($s$) apply. The superconductivity order parameters are
defined as ($\Delta_{\alpha ,\beta}>0$)
\begin{eqnarray}
\Delta_{\beta}=2W\sum_{\vec{k},\alpha}{}\sp{1,2}\langle
a_{\alpha\vec{k}\uparrow}a_{\alpha -\vec{k}\downarrow}\rangle \; ,\\
\nonumber \Delta_{\alpha}=2W\sum_{\vec{k}}\langle a_{\beta
-\vec{k}\downarrow}a_{\beta \vec{k}\uparrow}\rangle \; .
\end{eqnarray}

Here $W>0$ is the interband interaction constant. Its momentum
dependence is neglected for simplicity, as also the contribution
of intraband pairing channels. $W$ is supposed to contain
contributions of Coulomb (exchange) and electron-phonon nature
[26].

The diagonalization of (1) yields the gap equation ($\theta =k_BT$) system
\begin{eqnarray}
\Delta_{\beta}=W\Delta_{\alpha}\sum_{\vec{k},\alpha}{}\sp{1,2}
E_{\alpha}\sp{-1}(\vec{k}) th\frac{E_{\alpha}(\vec{k})}{2\Theta}\\ \nonumber
\Delta_{\alpha}=W\Delta_{\beta}\sum_{\vec{k}}
E_{\beta}\sp{-1}(\vec{k}) th\frac{E_{\beta}(\vec{k})}{2\Theta}
\end{eqnarray}
with the  quasiparticle energies
\begin{equation}
E_{\sigma}(\vec{k})=\pm\sqrt{\epsilon_{\sigma}\sp 2(\vec{k})
+\Delta_{\sigma}\sp 2(\vec{k})}\; .
\end{equation}
At $T_c$ the superconductivity gaps $\Delta_{\alpha ,\beta}$ tend
simultaneously to zero and one obtains from (5) an equation to
calculate $T_c$. The condensation energy ($H_{c0}\sp 2/8\pi$) is
represented by the thermodynamic critical field as
\begin{equation}
H_{c0}=\sqrt{4\pi [2\rho_{\alpha}\Delta\sp 2_{\alpha}+
\rho_{\beta}\Delta_{\beta}\sp 2]} \; .
\end{equation}

Inspection of the minimal quasiparticle excitation energies explains the
nature of the gaps expected to be observable in various doping windows [13].

\section{The excitations}
At heavy underdoping $c<c_0$ the manifestation of two pseudogaps is
expected in our model. The $\beta$-band connected pseudogap excitation
energy equals (in the superconducting state)
\begin{equation}
\Delta_{p\beta}=[(d_1-d-\alpha c\sp 2)\sp 2+\Delta\sp 2_{\beta}]\sp{1/2}
\end{equation}
being minimal for quasiparticles of this band. The corresponding
normal state gap ($\Delta_{\beta}=0$) closes at $c_0$. It
determines the energy separation of occupied and empty states,
being therefore an indicator of vanishing antiferromagnetic order.
The second pseudogap is larger and corresponds to the $\alpha 2$
band excitations in ($\frac{\pi}{2}, \frac{\pi}{2}$) spectral
window
\begin{equation}
\Delta_{p\alpha}=[d\sp 2+\Delta_{\alpha}\sp 2]\sp{1/2}\; .
\end{equation}
For $c\geq c_0$ the pseudogap $\Delta_{p\beta}$ vanishes
transforming into the $\beta$-band superconducting gap. It
determines now the minimal quasiparticle excitation energy of this
band. The smaller pseudogap continues at $c>c_0$ with stronger
doping dependence as
\begin{equation}
\Delta_{p\alpha}=[(\mu_2-\alpha c\sp 2)\sp 2+\Delta\sp 2_{\alpha}]\sp{1/2}\; .
\end{equation}
The superconducting gap $\Delta_{\alpha}$ can be discovered at this in the
($\pi ,0$) type window (the $\alpha 1$ band encloses $\mu$).

The normal state gap corresponding to $\Delta_{p\alpha}$ closes at
$c_x$. Beyond $c_x$ the excitation spectrum is expected to be
determined by both superconducting gaps $\Delta_{\beta}$ and
$\Delta_{\alpha}$. The latter must now be detectable in the whole
momentum window. With vanishing $\Delta_{p\alpha}$ at $c_x$ in the
normal state the tracks of AF ordering vanish at all. Note that
the pseudogaps appear in our model as precursors of
superconducting gaps on the doping, but not on the energetic
(phase diagram vertical) scale. The pseudogap vanishes, i.e.
transforms to superconducting gap in effectively or overdoped
samples as observed [4,6,19,20]. However the normal state gaps can
enter the superconducting energy domain before reaching zero. At
critical point where the pseudogap closes an insulator-metal
crossover is expected in the normal state. Such transition has
been observed experimentally [33]. And moreover -- the
characteristic resistivity temperature {\it vs} curve seems to
follow the pseudogap doping dependence.

\section{Illustrative results}
A seemingly plausible parameter set has been chosen to illustrate
the outcome of the presented model for a "typical'' electron-doped
cuprate superconductor. A maximal value of $T_c$ round 30 K near
$c=0.15$ in the superconductivity domain extending from $c=0.07$
until $c=0.3$ has given experimental guidance. The numerical
calculations have been made taking $d=0.03$; $d_1=0.1$; $\Delta
d=1.0$ and $\alpha =10$ (eV). The interband coupling constant
$W=0.175$ eV leads then to $T_{cm}=28$ K at $c=0.15$. The critical
electron doping concentrations read at this $c_0=0.08$ and
$c_x=0.13$.

In Figure 1 the dependence of transition temperature, chemical
potential and normal state (pseudo) gaps on doping are shown.
Remarkable values of $T_c$ are reached first near $c_0$ where two
overlapping bands enter into resonance with $\mu$. The maximal
$T_c$ is reached when the resonance of all three band components
have somewhat deepened beyond $c_x$. Concerning the absolute value
of $T_c$, the present model states that electron doping of the
strongly correlated system cannot proceed too far ($c<0.5$)
without a collapse of the basic spectrum. Extended doping leads to
quenching of the whole interband machinery. One sees in Fig.1 that
the smaller pseudogap can enter the superconductivity domain. For
the behavior of the pseudogaps cf. Fig.1 in [4] where the smaller
one is given under the question mark. The present model predicts
the presence of this second pseudogap.

In Figure 2 the calculated ($T=0$) superconductivity gaps and the
condensation energy ($H_{c0}$) on the doping scale are given.

Figure 3 illustrates the behavior of excitation gaps in the
superconducting state at $T=0$. One can see the transformation of
the larger pseudogap into the UHB $\beta$-superconducting gap
$\Delta_{\beta}$ at  $c_0$. For $c<c_0$ the contribution of
$\Delta_{\beta}$ is hidden in the $\Delta_{p\beta}$. The same
happens with the smaller pseudogap and the defect superconducting
gap $\Delta_{\alpha}$ in the ($\frac{\pi}{2}, \frac{\pi}{2}$)
window. Note that at $c<c_x$ the gap $\Delta_{\alpha}$ (see the
fragment of it in Fig.2) must be detectable in the ($\pi ,0$)
window. At $c>c_x$ $\Delta_{\alpha}$ must be observable in both
windows as seen in Fig.2. The energy scales of the pseudo- and
superconducting gaps can be comparable out of the underdoped
region. This can prepare difficulties when interpreting fragmental
data of measurements. In analyzing the manifestations of two
spectral gaps there can arise problems because the states
associated with the smaller gap can mask the sharpness of the
larger gap. The peak associated with the latter can then appear as
a spectral hump, cf. [6]. The characteristic gap relations
$2\Delta_{\alpha ,\beta}/kT_c$ violate the BCS universality and
does not change markedly with doping being about 4.8 for the
$\beta$ and 2.6 for the $\alpha$ band. The condensation energy
given also in Fig.2, shows the usual bell-like dependence.

As a result it seems that the present model is able to explain the
gross nature of the electron-doped cuprate phase diagram. Detailed
investigations of the number of pseudogaps and on the
superconducting gaps in the whole doping region in a single
experiment were of considerable interest to prove some of our
conclusions looking as predictions.

At present one can seemingly constatate that in both cases of hole
and electron doped cuprate superconductors the deepest background
is the same. The necessary doping metallizes the basic strongly
correlated materials and prepares transformed electron structures
with new energetic states. However, these novel backgrounds having
comparable elements and functioning are realized in different
manner and on different sublattices. In the case of electron
doping it is the Cu-sublattice and doping influences immediately
the spectrum built by strong electron correlations. In the case of
hole doping it is the highly polarizable and deformable oxygen
sublattice. The universal multiband pairing mechanism leads in
both cases to the comparable characteristic phase diagrams of
high-temperature superconductivity.

This work was supported by Estonian Science Foundation grant No
6540. The authors are indebted to T.\"Ord for discussions.

\newpage

Fig.1. The transition temperature (1), chemical potential (2), the
large (3) and small (4) normal state (pseudo)gaps of a cuprate on
the electron doping (c) scale.

Fig.2. The calculated zero temperature superconducting gaps and
the thermodynamic critical field representing the condensation
energy $H_{co}\sp 2 (8\pi )\sp{-1}$.

Fig. 3. The excitation energies represented by the pseudogap and
superconducting components on the electron doping scale
 ($T =0$). 1 -- $\Delta_{p\alpha}$ and $\Delta_{\alpha}$;
 2 -- $\Delta_{p\beta}$ and $\Delta_{\beta}$.
\newpage


%


\begin{thebibliography}{}
\bibitem{1} A.Damascelli, Z.Hussain, Z.-X-Shen, Rev. Mod. Phys.
{\bf 75}, (2003) 473.
\bibitem{2} N.Harima et al., Phys. Rev. B {\bf 64}, (2001) 220507R.
\bibitem{3} A.Biswas et al., Phys. Rev. B {\bf 64 }, (2001) 104519.
\bibitem{4} L.Alff et al., Nature {\bf 422}, (2003) 698.
\bibitem{5} T.Tohyama, Physica C {\bf 412-414}, (2004) 139.
\bibitem{6} H.Matsui et al., Phys. Rev. Lett. {\bf 94}, (2005) 047005.
\bibitem{7} S.R.Park et al., Phys. Rev. B {\bf 75}, (2007) 060501R.
\bibitem{8} S.Uchida et al., Phys. Rev. {\bf 43} (1991) 7942.
\bibitem{9} M.Tachiki, in: {\it Strong Correlation and Superconductivity},
edited by H.Fukuyama et al. (Springer, Berlin, 1989) p. 139.
\bibitem{10} T.Takahashi, Physica C {\bf 170}, (1990) 416.
\bibitem{11} M.B.J.Meinders, H.Eskes, G.A.Sawatzky, Phys. Rev. {\bf 48},
(1993) 3916.
\bibitem{12} N.Kristoffel, P.Rubin, in {\it Symmetry and Heterogeneity in
High-Temperature Superconductors}, edited by A.Bianconi (Springer, 2006) p.55.
\bibitem{13} N.Kristoffel, P.Rubin, Solid State Commun. {\bf 122}, (2002) 265.
\bibitem{14} {\it Phase separation in cuprate superconductors}, edited by
K.A.M\"uller and G.Benedek (World Sci., Singapore 1993).
\bibitem{15} C.Kusko et al., Phys. Rev. B {\bf 66}, (2002) 140513(R).
\bibitem{16} Q.Yuan, X.-Z.Yan, C.S.Ting, Phys. Rev. B {\bf 74}, (2006)
215403.
\bibitem{17} B.Kyung et al., Phys. Rev. Lett. {\bf 93}, (2004) 147004.
\bibitem{18} D.K.Sunko, S.Barisic, Phys. Rev. B {\bf 75}, (2007) 060506(R).
\bibitem{19} N.P.Armitage, Phys. Rev. Lett. {\bf 87}, (2001) 147003.
\bibitem{20} Y.Onose et al., Phys. Rev. Lett. {\bf 87}, (2001) 217001.
\bibitem{21} N.Kristoffel, P.Rubin, Physica C {\bf 402}, (2004) 257.
\bibitem{22} N.Kristoffel, P.Rubin, J. Supercond. {\bf 18}, (2005) 705.
\bibitem{23} R.Khasanov et al., Phys. Rev. Lett. {\bf 98}, (2007) 057007.
\bibitem{24} H.Suhl, B.T.Matthias, L.R.Walker, Phys. Rev. Lett. {\bf 3},
(1959) 552.
\bibitem{25} V.A.Moskalenko, Fiz. Met. Metalloved. {\bf 8}, (1959) 503.
\bibitem{26} N.Kristoffel, P.Konsin, T.\"Ord, Rivista Nuovo Cim. {\bf 17},
(1994) 17.
\bibitem{27} A.Bianconi, J. Supercond. {\bf 18}, (2005) 25.
\bibitem{28} A.Bianconi, M.Filippi, in: {\it Symmetry and Heterogeneity in
High-Temperature Superconductors}, edited by A.Bianconi (Springer, 2006) p. 21.
\bibitem{29} N.Kristoffel, T.\"Ord, P.Rubin, Physica C {\bf 437-438}, (2006)
168.
\bibitem{30} N.Kristoffel, P.Rubin, Phys. Lett. A {\bf 356}, (2006) 249;
ibid. {\bf 360}, (2006) 367.
\bibitem{31} N.Kristoffel, T.\"Ord, P.Rubin, in {\it Electron Correlation
in New Materials and Nanosystems}, edited by K.Scharnberg and S.Kruchinin
(Springer, 2007) p. 275.
\bibitem{32} K.A.M\"uller, Physica C {\bf 341-348}, (2000) 11.
\bibitem{33} P.Fournier et al. Phys. Rev. Lett. {\bf 81}, (1998) 4720.
\end{thebibliography}
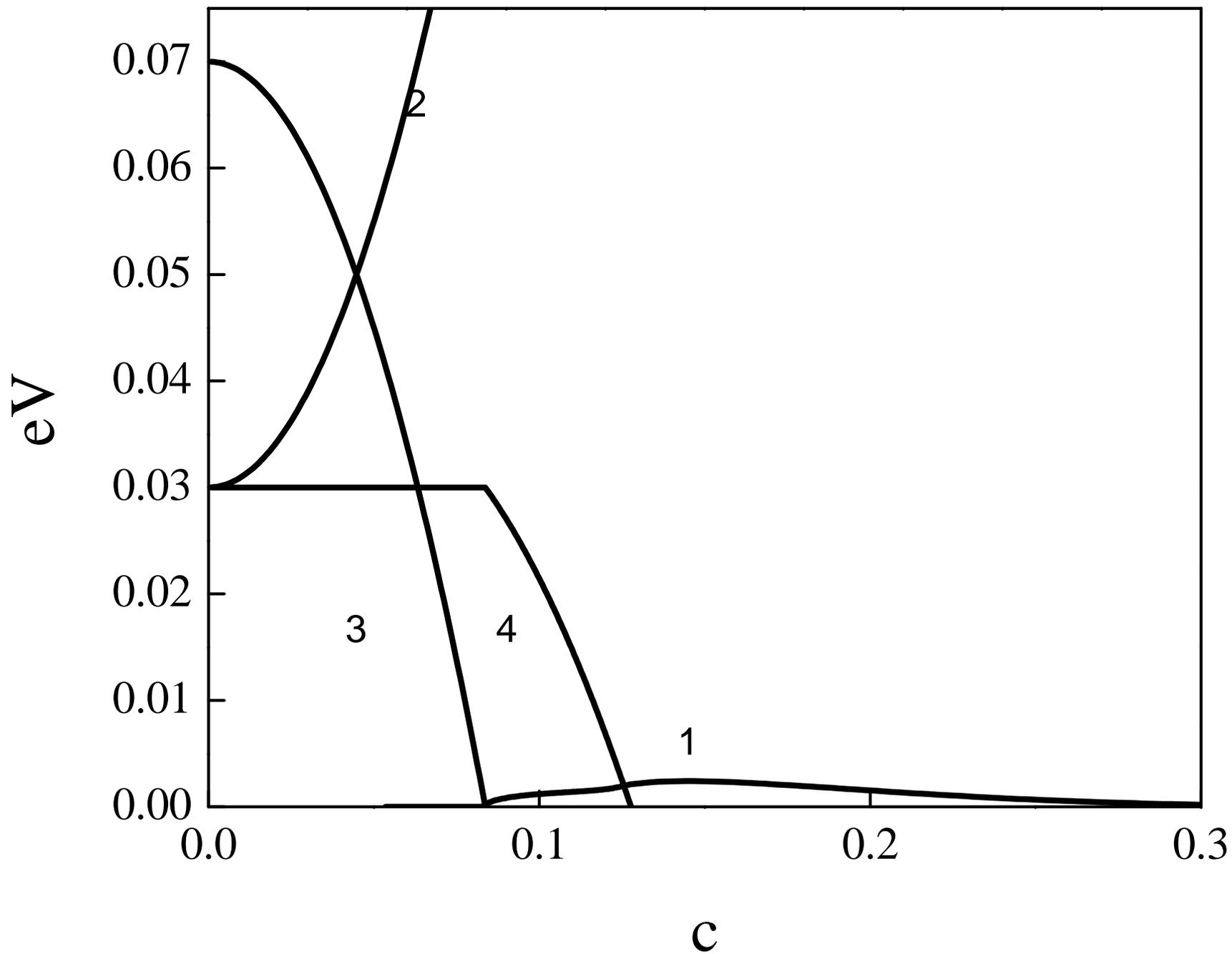
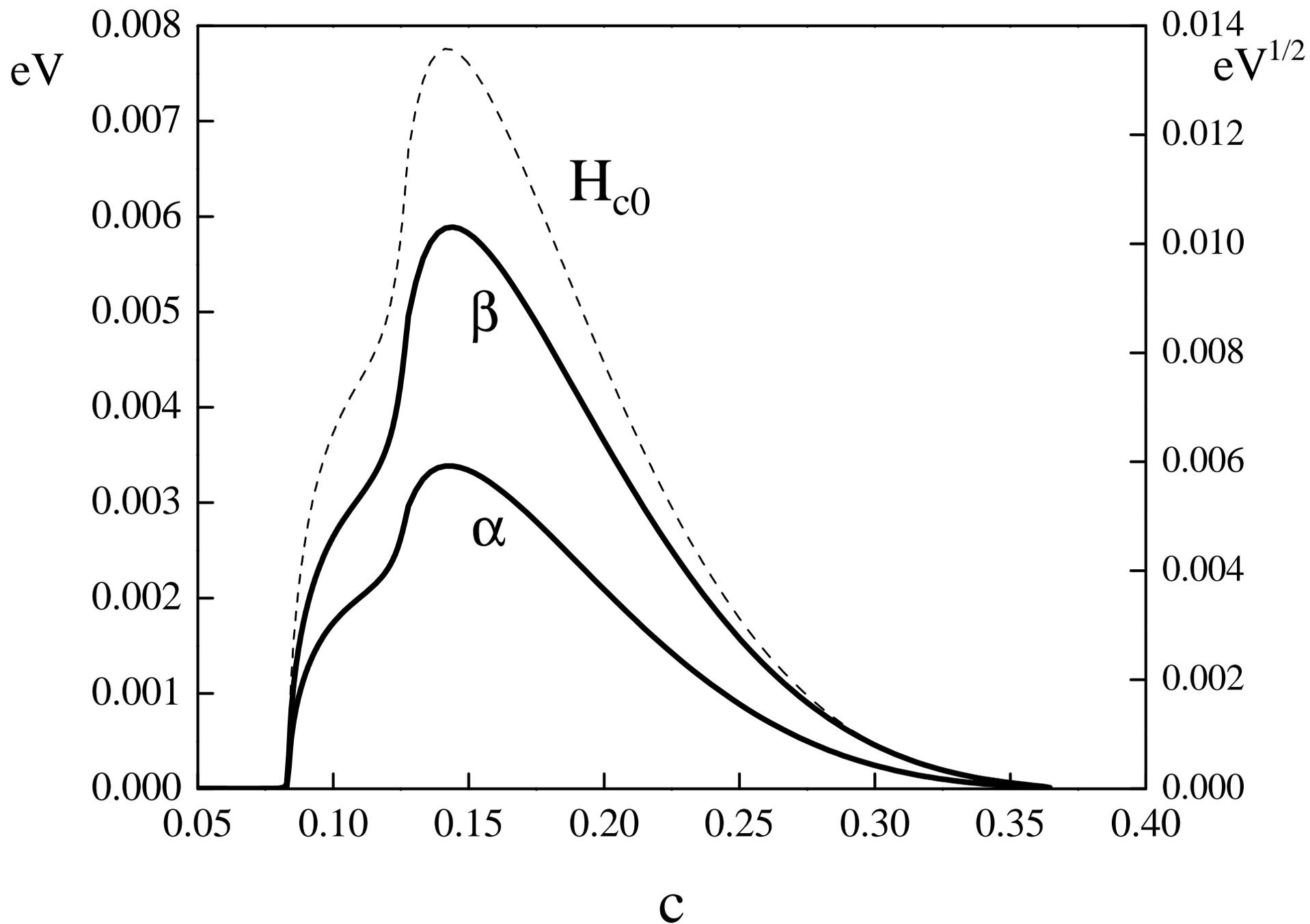
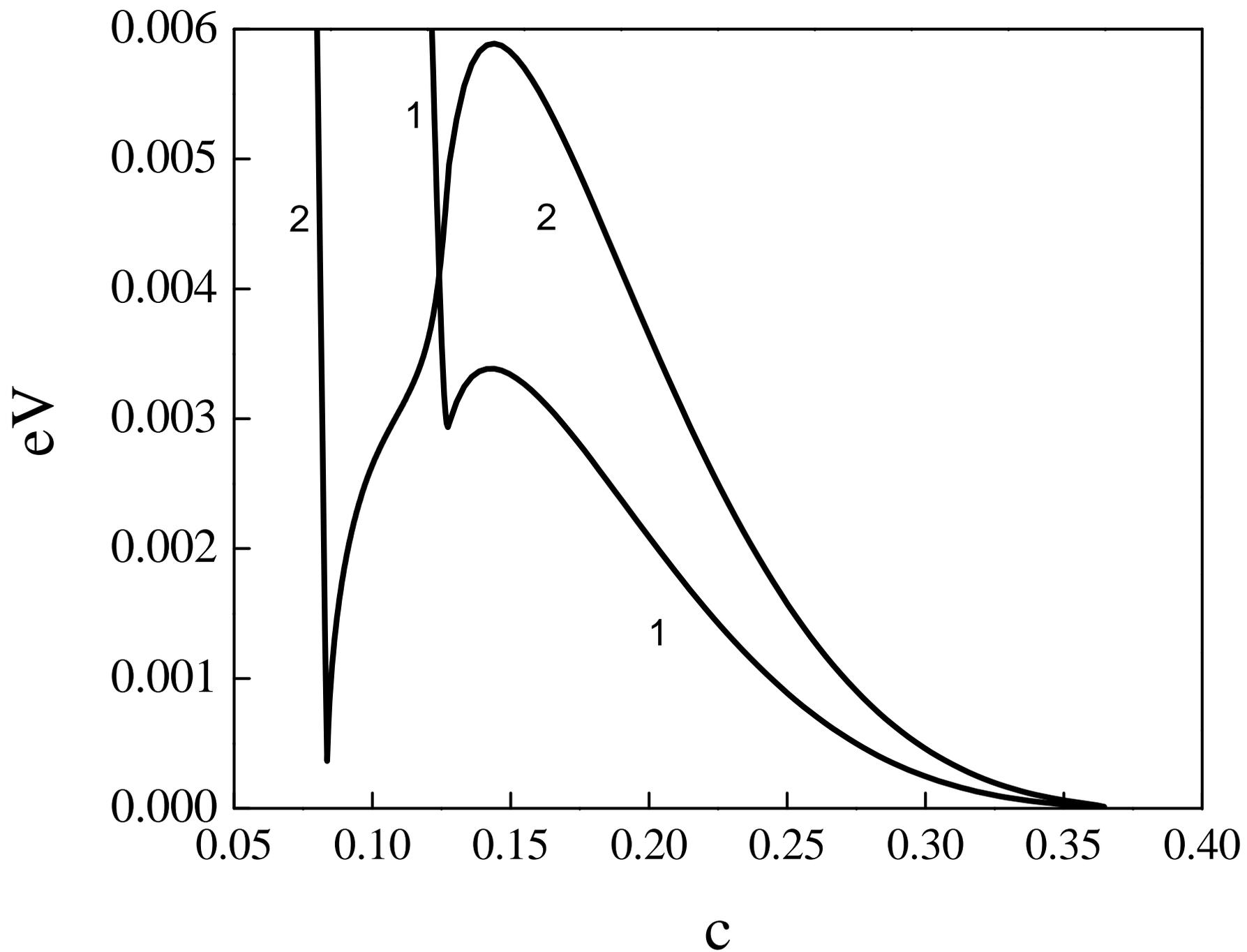
\end{document}